\documentclass[12pt]{iopart}

\usepackage{iopams}
\usepackage{bm}
\usepackage{color}
\usepackage{graphicx}

\begin{document}

\title[Density down-ramp injection subject to a magnetic field]{Controlling injection using a magnetic field to produce sub-femtosecond bunches in the laser wakefield accelerator}

\author{Q. Zhao$^{1,2}$, S. M. Weng$^{1,2,\dagger}$, M. Chen$^{1,2}$, M. Zeng$^{3}$, B. Hidding$^{4,5}$,D. A. Jaroszynski$^{4,5}$, R. Assmann$^3$, Z. M. Sheng$^{1,2,4,5,\ddagger}$}

\address{$^1$ Key Laboratory for Laser Plasmas (MoE), Department of Physics and Astronomy, Shanghai Jiao Tong University, Shanghai 200240, China}%
\address{$^2$ Collaborative Innovation Center of IFSA, Shanghai Jiao Tong University, Shanghai 200240, China}%
\address{$^3$ Deutsches Elektronen-Synchrotron DESY, 22607 Hamburg, Germany}
\address{$^4$ SUPA, Department of Physics, University of Strathclyde, Glasgow G4 0NG, UK}
\address{$^5$ Cockcroft Institute, Sci-Tech Daresbury,
Cheshire WA4 4AD, UK}

\ead{\mailto{$^\dagger$wengsuming@gmail.com}, or \mailto{$^\ddagger$z.sheng@strath.ac.uk}}%

\begin{abstract}
It is shown that electron injection into a laser-driven plasma bubble can be manipulated by applying an external magnetic field in the presence of a plasma density gradient. The down-ramp of the density-tailored plasma locally reduces the plasma  wave phase velocity, which triggers injection. The longitudinal magnetic field dynamically induces an expanding hole in the electron density distribution at the rear of the wake bubble, which reduces the peak electron velocity in its vicinity.
Electron injection is suppressed when the electron velocity drops below the phase velocity, which depends on the size of the density hole. This enables the start and end of electron injection to be independently controlled, which allows generation of sub-femtosecond electron bunches with peak currents of a few kilo-Ampere, for an applied magnetic field of $\sim 10$ Tesla.
\end{abstract}


%
%
\submitto{\PPCF}
%
%


\section{Introduction}
A plasma density wave or wake driven by the ponderomotive force of
an ultraintense laser pulse can trap electrons and accelerate them
to high energies \cite{Tajima1979,Esarey2009}. This well-known
laser wakefield acceleration (LWFA) promises compact sources of
high energy electrons because of the ultra-strong
accelerating electric fields, which can exceed $100$ GV/m, in the so-called bubble
regime that is characterized by a spherical electron cavity containing ions and surrounded by
a high-density electron sheath \cite{Pukhov2002,Lu2006,Yi2013}.
In particular, the electron bunches obtained in the
LWFA can be ultrashort, which is a major advantage of the LWFA and
of great significance as drivers of ultrashort X-ray sources and
potential compact X-ray free-electron lasers
\cite{Jaroszynski2002,Rousse2004,Schlenvoigt2008,Fuchs2009,Cipiccia2011,Ersfeld2014,Chen2016}.
In general, the high-quality electron bunch should reside within the accelerating and focusing region of a wakefield with a length about $\lambda_p/4$ in the linear regime \cite{Faure2006,Lundh2011}, where $\lambda_p$ is the relativistic plasma wavelength.
It is therefore expected that ultrashort electron bunches generated by a typical LWFA will have durations $\lambda_p/4c \sim 10$ femtoseconds for plasma densities $n_e \simeq 10^{18}\sim10^{19}$ cm$^{-3}$.
There is considerable interest in generating even
shorter electron bunches, with attosecond durations, for various
applications including attosecond X-ray sources and direct imaging
\cite{Kartner2016,Dorda2016,Morimoto2018,Hassan2018}.

In LWFA, a shorter bunch duration down to sub-femtosecond is
possible if the injection of electrons into the wake is
highly localized. The localized electron injection can be achieved
by near-threshold self-injection
\cite{Buck2011,Heigoldt2015,Islam2015}, by colliding pulse
injection \cite{Faure2006,Lundh2011}, or by up-ramp density
transition \cite{Li2015}. Alternatively, the localized electron
injection can be realized by manipulating the plasma wake
structure. For instance, the local plasma wake phase and wavelength can be tuned by longitudinal plasma density tailoring \cite{Bulanov1998}. To facilitate electron injection,
the wake phase velocity can be reduced momentarily by a
density down-ramp
\cite{Geddes2008,Gonsalves2011,Hansson2015,Martinez2017,Xu2017,Tooley2017}. In
particular, the highly localized injection can be achieved if the
time in which the peak forward-directed plasma electron velocity exceeds the wake phase
velocity is ultrashort, which promises the generation of
attosecond electron bunches \cite{Tooley2017}. So far, the
generation of isolated sub-femtosecond electron bunches has not
yet been demonstrated experimentally and several technical
challenges still need to be overcome.

It is well known that the wakefield structure in the LWFA can be modified by a static external magnetic field. This may provide an alternative approach to control the electron injection \cite{Hosokai2006,Vieira2011,Bulanov2013,Rassou2015,Zhao2018}.
By imposing an external transverse magnetic field that is on the order of hundreds of Tesla, the longitudinal trapping condition in the self-injection regime can be significantly relaxed \cite{Vieira2011}, which also enhances the charge number of injected electrons.
In contrast, it has been recently found that the external magnetic field required to modify the transverse trapping condition in the ionization-injection regime is only on the order of tens of Tesla \cite{Zhao2018}, which promises efficient generation of high-quality electron bunches with both high charge and low energy spread.
It is noted that strong magnetic fields of a few tens of Tesla can be generated in a small volume by either traditional technology in laboratories \cite{Pollock2006,Fiksel2015} or a novel proposal using twisted laser beams \cite{Shi2018}.

In this paper, we present a study of the
manipulation of the laser-driven plasma bubble to control the persistence of  electron injection, through combining a density-profile-tailored plasma with a longitudinal magnetic field.
We show that the static longitudinal magnetic field modifies the transverse structure of
the bubble, while the density gradient changes its longitudinal
structure. The magnetic field induces a radial density hole in the
bubble rear \cite{Bulanov2013,Rassou2015}, which expands and as the bubble evolves along the density down-ramp. Electron injection is triggered
by the decreasing phase velocity of the bubble along the density
down-ramp, and then suppressed by the expanding hole at the bubble
rear. In this way, the position and persistence of electron injection can
be controlled, leading to injection of isolated sub-femtosecond electron bunches.

\section{Theoretical analysis}
We first consider the effect of the plasma density gradient on the
laser-driven plasma wake. In a tenuous inhomogeneous plasma, the
wake wave has a local phase $\chi=k_p(z)\xi$, where $\xi=z-v_pt$ is the relative coordinate, $\omega_p=(n_ee^2/\varepsilon_0m_e)^{1/2}$ and
$k_p=\omega_p/v_p$ are the local plasma frequency and wavenumber,
respectively. For a non-relativistic laser pulse, the wavelength
$\lambda_{p}(z)$ only depends on the local plasma density
$n_e(z)$, and therefore $\beta_{p}=\beta_g (1+
(\chi/2\pi)(d\lambda_{p}/dz))^{-1}$, where $\beta_{p}=v_p/c$ and
$\beta_{g}=v_g/c$ are the normalized wake phase and laser group
velocities, respectively. For a relativistic laser pulse with initial peak normalized potential amplitude $a_0\gg1$, the effect of laser amplitude evolution on
$k_p(z)$ and $\omega_p(z)$ must be considered appropriately. In
this case, the ponderomotively expelled electrons oscillate
transversely at the relativistic betatron frequency
$\omega_\beta=\omega_p/\sqrt{2\gamma_e}$ with the Lorentz factor
$\gamma_e \simeq \sqrt{1+a^2/2}$ in the ponderomotive approximation
\cite{Tsakiris2000}. The initially stationary electrons return to
the laser axis after half a betatron oscillation period
$\pi/\omega_\beta$ and cross at the bubble rear. The velocity of the
bubble rear, or the phase velocity at $\chi=-2\pi$, where the longitudinal electric field $E_z=0$, can be
formulated as $\beta_p^{-1}=\beta_g^{-1}+cd\tau/dz$
\cite{Tooley2017}, where $\tau=2\pi/\omega_\beta$. This
approximately gives the bubble velocity as
\begin{equation}
\beta_{p}=\beta_g \left[1 - \beta_g \frac{\lambda_{p0}}{4} \left( \frac{\sqrt{\gamma_e}}{\tilde n^{3/2}}\frac{d \tilde n}{dz}-\frac{1}{\sqrt{\tilde n\gamma_e}}\frac{d\gamma_e}{dz}\right)\right]^{-1}\label{ph.vel.},
\end{equation}
where $\tilde{n}=n_e/n_0$ is the plasma density normalized to the
plateau density $n_0$, and $\lambda_{p0}$ is the plasma wavelength
corresponding to $n_0$. The effects of the plasma density gradient and
laser amplitude evolution are included in the terms $d \tilde
n/dz$ and $d\Gamma_e/dz$, respectively.  The bubble velocity
can be reduced by decreasing the plasma density and/or increasing the
laser amplitude.

Under a strong longitudinal magnetic field $B_0$, however, the bubble rear will open up \cite{Bulanov2013,Rassou2015}.
In this case, the electrons moving along the bubble sheath experience a time-varying magnetic field  $\phi=B_0\pi r_b^2(\xi)$, where $r_b$ is the radius of the bubble. This induces an azimuthal electric field, which causes the sheath electrons to obtain an azimuthal velocity $v_\varphi$ and rotate reversely around the laser axis.
As a consequence, a huge longitudinal magnetic field $B_z$ is self-generated and distributes locally inside the density-hole region, with the same direction as  $B_0$, as described by Lenz's law \cite{Rassou2015}. Finally, the radial motion of sheath electrons is governed by
\begin{equation}
\frac{\partial p_r}{\partial t}+v_r\frac{\partial p_r}{\partial r}=\frac{p_\varphi v_\varphi}{r}-eW_\perp-ev_\varphi(B_0+B_z), \label{motion}
\end{equation}
where $p_\varphi v_\varphi/r$ is the centrifugal force, and
$W_\perp=E_r-cB_\varphi$ is the radial wakefield
\cite{Lu2006,Yi2013}. It is this centrifugal force that opens up the bubble rear, because  it tends to infinity when $r\rightarrow 0$.
Considering the equations for the transverse momentum only, the radius of the hole in the opened bubble rear should be governed mainly by the plasma density and magnetic field, which can be approximated by $r_{\rm{min}}\approx2\sqrt{2}(c\omega_{c})/\omega_p^2$ \cite{Bulanov2013}, where  $\omega_{c}=eB_0/m_e$ is the electron cyclotron frequency. Nevertheless, since electrons experience a strong longitudinal acceleration in the wakefield, the hole radius also depends on the instantaneous mean electron energy around the bubble rear, because the centrifugal force $p_\varphi v_\varphi/ r = m_e \gamma_ev^2_\varphi/r \propto\gamma_e$.

\section{PIC simulations}

To demonstrate the combined effects of plasma density tailoring and the
longitudinal magnetic field on the wake structure, and therefore the
consequent electron injection, three-dimensional (3D)
particle-in-cell (PIC) simulations using OSIRIS \cite{Fonseca} have
been performed. In each simulation, a $30\times30 \times 70$
$\mu$m$^3$ simulation box moves along the z-axis at the speed of
light. It is subdivided  into $120\times 120\times 2240$ cells
with $1\times1\times 2$ particles per cell. The plasma comprises a
plateau background density $n_0$ corresponding to a plasma frequency $\omega_{p0}$ and a Gaussian density bump
\begin{equation}
\frac{n_e}{n_0}=1+\alpha\exp(-(z-z_i)^2/2\sigma_z^2),
\end{equation}
where $\alpha$ is the relative amplitude of the density peak at $z_i$, $\sigma_z$ is the characteristic length of the Gaussian bump. Such a
plasma density profile is realizable in experiments as longs as the ramp does not need to be too steep \cite{Hansson2015,Kononenko2016}. A linearly polarized (along y-direction) laser pulse with duration of 30 fs, with an initial peak normalized amplitude $a_0=4$ and a waist of 15 $\mu$m at focal plane $z=0$, is used. The laser wavelength is 0.8 $\mu$m, correspondingly the critical plasma density $n_c=1.7\times10^{21}$ cm$^{-3}$. A uniform external magnetic field $B_0$ is assumed to be along the $z$-axis and is exerted on the whole plasma region.

\subsection{Evolution of phase velocity due to bubble stretch}

\begin{figure}
\centering\includegraphics[width=0.6 \textwidth]{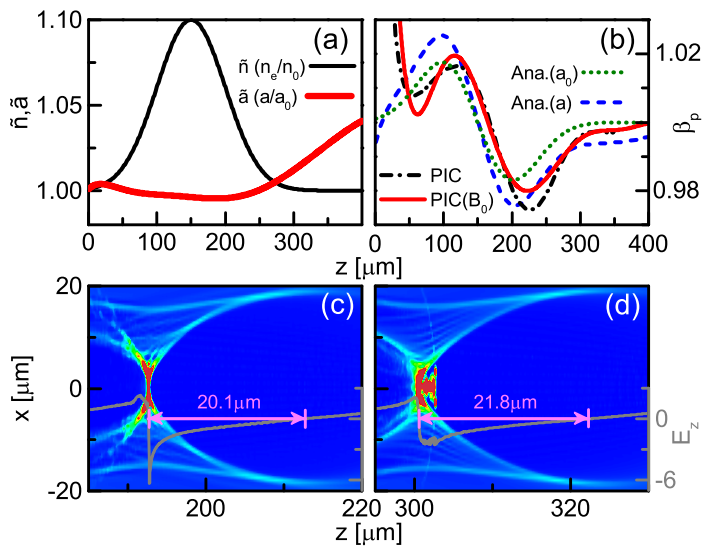}
\caption{(Colour online) (a) Tailored plasma density profile with down-ramp parameters: $n_0=0.0005n_c$, $z_i=150$ $\mu\rm{m}$, $\alpha=0.1$, and $\sigma_z=50$ $\mu\rm{m}$. In-situ laser amplitude obtained from the simulation, where $a_0=4$.
(b) Bubble velocities obtained from Eq. (\ref{ph.vel.}) using $n_e$ with constant $a_0$ (dot curve) and evolving $a$ (dashed curve) as shown in (a), and the simulations with $B_0=0$ (dash-dot curve) and $B_0=50$ $\rm{T}$ (solid curve), respectively.
(c) and (d) Plasma bubble structures, superimposed on the on-axis $E_z$ (normalized to $m_ec\omega_{p0}/e$), when the bubble rear locates at the slope ($z_b=192.7$ $\mu\rm{m}$) and the bottom ($z_b=300.7$ $\mu\rm{m}$) of the down-ramp, respectively.
} \label{phase vel.}
\end{figure}

Figure \ref{phase vel.}(a) displays the local plasma density and
the laser amplitude as functions of the propagation
distance $z$. It is seen that the laser amplitude decreases slightly along almost the whole density bump region before z=200 $\mu$m and then increases rapidly because of self-focusing. Substituting into Eq. (\ref{ph.vel.}), the
local wake phase velocity can be calculated analytically. Figure
\ref{phase vel.}(b) illustrates that the bubble rear velocity
decreases dramatically due to the decreasing plasma density at the
down-ramp around $z \sim 200$ $\mu$m, and the simulation result is
in good agreement with the analytical result with evolving laser amplitude $a$, which is compared with the one with constant $a_0$. The reduction in the
bubble velocity is attributed to the increase of the wavelength of
the wake. The longitudinal stretch of the wake bubble is confirmed
in figures \ref{phase vel.}(c) and \ref{phase vel.}(d), where the
radius of the bubble increases from $20.1$ to $21.8$ $\mu$m.
As a consequence, electron injection is triggered, as is illustrated
in figure \ref{phase vel.}(d). From figure \ref{phase vel.}(b), one
notes that the reduction in the bubble velocity is slightly
weakened by a longitudinal magnetic field $B_0=50$ T.
This is because the continuously injected electron charge in the $B_0=0$
case slightly pushes back the bubble rear, due to the strong Coulomb
repulsion, and hence reduces the bubble velocity further, while suppressing  electron injection for non-zero $B_0$, as will be discussed below. Because both the laser evolution and the injected charge can reduce the phase velocity, we observe that it is hard to finely control the density-gradient profile to produce stable sub-femtosecond electron bunches.

\subsection{Magnetic field induced injection suppression along a density down-ramp}
\begin{figure}
\centering\includegraphics[width=0.6 \textwidth]{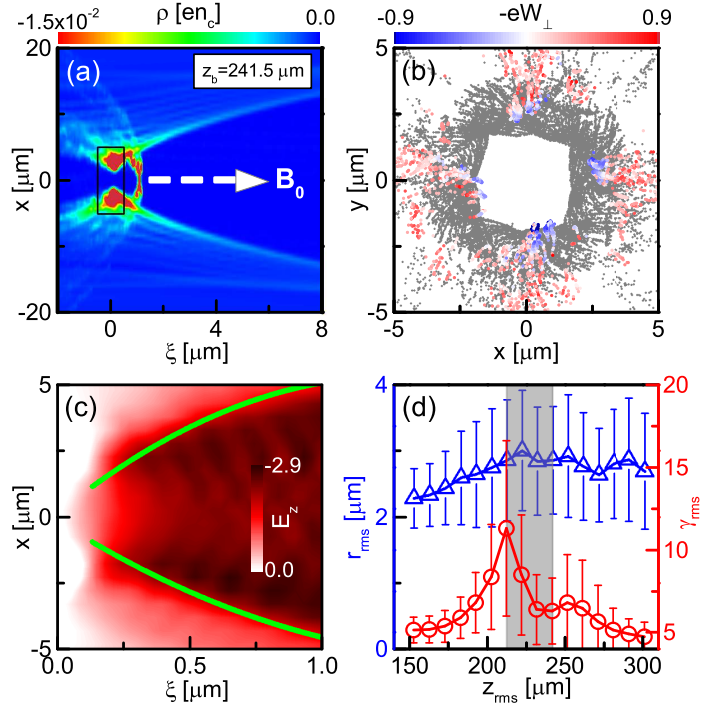}
\caption{(Colour online) (a) Bubble structure for the bubble rear
located at $z_b=214.5$ $\mu\rm{m}$. (b) Transverse distribution of
the electrons in the rectangle region in (a), the electrons with a
longitudinal velocity exceeding the local bubble velocity,  coloured
according to their experienced force of transverse wakefields
$-eW_\perp$ (normalized to $m_ec\omega_{p0}$). (c) $E_z$
corresponding to the bubble in (a). Green lines outline the
bubble sheath. (d) Time evolution of the root-mean-square (RMS) radius $r_{\rm{rms}}$
and the Lorentz factor $\gamma_{\rm{rms}}$ of energetic electrons
around the bubble rear, where their RMS distributions are given. Grey
shaded region marks the space where electron injection occurs. Except for
$B_0=50$ T, all other parameters are the same as those in figure
\ref{phase vel.}. } \label{hole}
\end{figure}

To understand the effect of a non-zero $B_0$ on the electron
injection, it is important to reveal first its effect on the
bubble structure. As previously predicted
\cite{Bulanov2013,Rassou2015}, figure \ref{hole}(a) confirms the
appearance of an open bubble rear for $B_0=50$ T. More
importantly, the electron injection in this case is found to take
place over a short distance. The electron bunch ($\sim0.5$ $\mu$m)
resulting from this highly localized injection is much shorter
than that ($\sim2$ $\mu$m) in the case with $B_0=0$. The transverse
distribution of the electrons around the open bubble rear is
displayed in figure \ref{hole}(b), where the suppression of electron injection is confirmed because only a small number of electrons around the bubble rear can simultaneously achieve the local bubble velocity and be focused with $-eW_\perp<0$.
Because of the open bubble rear, the wake
accelerating field $E_z$ has its maximum amplitude away from the laser
axis, as shown in figure \ref{hole} (c). As a consequence, the most
efficient acceleration region for the electrons moving along the
bubble sheath is located at a distance away from the axis.
Defining $\gamma_{\rm{rms}}$, $r_{\rm{rms}}$ and $z_{\rm{rms}}$ as
the RMS Lorentz factor, radius and longitudinal
position of energetic electrons ($\gamma>4$) around the bubble rear,
figure \ref{hole}(d) shows that both $\gamma_{\rm{rms}}$ and
$r_{\rm{rms}}$ increase before the occurrence of electron
injection. The increase of $\gamma_{\rm{rms}}$ can be attributed
to the prolonged acceleration time in the expanding bubble along the
density down-ramp. With increasing $\gamma_{\rm{rms}}$,
however, $r_{\rm{rms}}$ also increases because the centrifugal
force $p_\varphi v_\varphi/ r \propto \gamma_{\rm{rms}}$. As a result, finally, the increase of $r_{\rm{rms}}$ inhibits electron injection.

\begin{figure}
\centering\includegraphics[width=0.6 \textwidth]{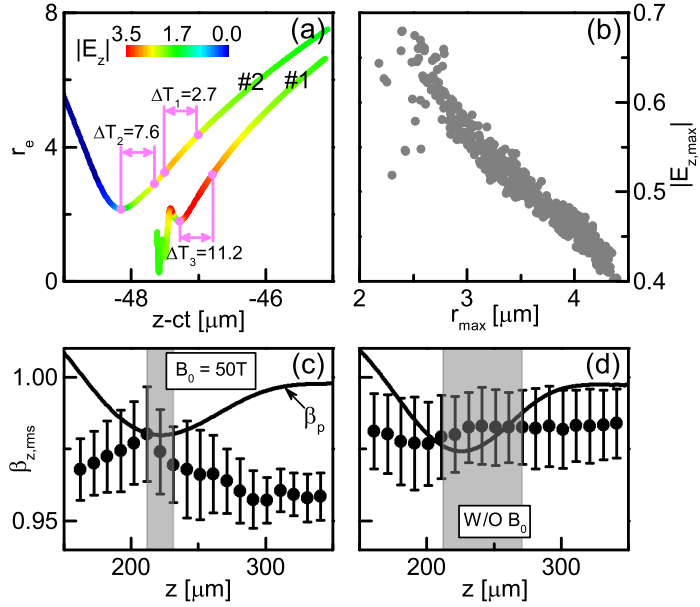}
\caption{(Colour online) (a) Typical electron trajectories selected
from the injected bunch ($\#1$) and the unclosed bubble rear
($\#2$) in figure \ref{hole}(a), respectively. Different standing
time in some typical acceleration regions (with the same
$\Delta(z-ct)=0.5$ $\rm{\mu m}$) are marked as $\Delta T_{1,2,3}$. (b)
Correlation between the maximums of $E_z$ and the radial positions
where these maximums are achieved for 500 electrons randomly
selected from figure \ref{hole}(b). The RMS longitudinal velocity $\beta_{z,rms}$ with its RMS distribution for the
energetic electrons at the bubble rear vs the simulated local bubble velocity $\beta_p$ along the density down-ramp  with (c) and  without (d) $B_0=50$ T, respectively. The grey shaded regions  mark the injection regions.} \label{track}
\end{figure}

To explain the suppression of electron injection through increasing
$r_{\rm{rms}}$, the instantaneous accelerating fields $E_z$ along
the trajectories of two typical electrons are compared in figure
\ref{track}(a). Firstly, we find that the injected electron $\# 1$ that is
closer to the axis experiences a stronger accelerating field than
the non-injected electron $\# 2$. Defining $|E_{\rm{z,max}}|$ and
$r_{\rm{max}}$ as  the maximum of $E_z$ and the corresponding
radial position where this maximum is achieved, figure
\ref{track}(b) shows a linear negative correlation between
$|E_{\rm{z,max}}|$ and $r_{\rm{max}}$. Secondly, we find that the
most efficient acceleration region for electron $\# 1$ is much
closer to its turning point than that for electron $\# 2$.
Therefore, electron $\# 1$ can stay in the efficient
acceleration region for a longer period of time than electron
$\# 2$. The stronger accelerating field and longer accelerating
time combine to guarantee that electron $\# 1$ can be
accelerated to the bubble velocity and therefore be trapped. In
contrast,  electron $\# 2$ can not be injected since it is
further away from the axis. With increasing $r_{\rm{rms}}$,
more and more energetic electrons will be far away from the axis,
as  electron $\# 2$. As a result, electron injection is
inhibited for these electrons. Figure \ref{track}(c) shows that electron injection
is triggered as soon as the increasing RMS velocity
$\beta_{\rm{z,rms}}$ of energetic electrons exceeds the decreasing wake
phase velocity along the density down-ramp. However, injection ends promptly because $\beta_z$ quickly decreases due to the
increasing $r_{\rm{rms}}$ for $B_0=50$ T. In contrast, figure
\ref{track}(d) shows that electron injection lasts a longer
time since $\beta_z$ is nearly constant under $B_0=0$.
\subsection{Subfemtosecond electron bunches produced by 3D manipulation of the plasma bubble}
\begin{figure}
\centering\includegraphics[width=0.6 \textwidth]{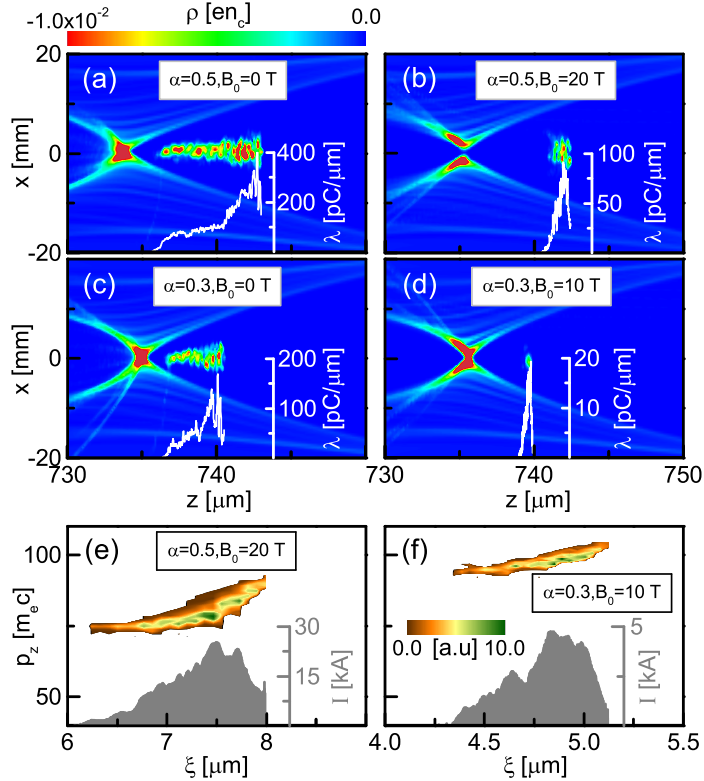}
\caption{(Colour online)
Bubble structures and injected electron bunches in density-profile-tailored plasma with $n_0=0.0002n_c$, $\sigma_z=100$ and (a) $B_0=0$, $\alpha=0.5$; (b) $B_0=20$ T, $\alpha=0.5$; (c) $B_0=0$, $\alpha=0.3$; (d) $B_0=10$ T, $\alpha=0.3$. Insets: Charge profiles of the injected electron bunches. (e)-(f) Phase-space distributions and currents of the injected electron bunches in (b) and (d), respectively.} \label{bunch}
\end{figure}

From the above analysis, it is evident that electron injection
can be flexibly controlled by the combination of a density
down-ramp and a magnetic field $B_0$. The density down-ramp
triggers the electron injection, while the magnetic field $B_0$ suppresses
the injection. The required $B_0$ to suppress the injection decreases
with decreasing plasma density $n_0$ because the radius of the
hole in the open bubble rear is inversely proportional to
$n_0$ \cite{Bulanov2013}. Figures \ref{bunch}(a) and
\ref{bunch}(b) illustrate that a weaker magnetic field $B_0=20$ $\rm{T}$
can suppress injection for a lower density $n_0=0.0002
n_c$, where the corresponding length of injected electron bunch is
reduced from $\sim8$ to $\sim1.5$ $\mu\rm{m}$. The required $B_0$
for suppressing injection can be further reduced by reducing the plasma density gradient. Figures \ref{bunch}(c) and
\ref{bunch}(d) illustrate that electron injection can be
effectively suppressed by $B_0=10$ T if a more gentle down-ramp is
adopted ($\alpha=0.3$), which is also more feasible experimentally. Figures \ref{bunch}(e) and \ref{bunch}(f) show the phase-space distributions and currents of the ultrashort
electron bunches presented in figures \ref{bunch}(b) and
\ref{bunch}(d), respectively. These quasi-monoenergetic electron
bunches have charges of $35(7)$ pC and RMS durations of
$\tau_{\rm{rms}}=1.291(0.594)$ fs, which corresponds to peak currents
$23.2(5)$ kA, respectively. Their emittances are as low as
$\varepsilon_{n,x(y)}=1.2(1.3)$ $\mu$m and
$\varepsilon_{n,x(y)}=0.8(1.1)$ $\mu$m, respectively.

\section{Conclusion}
In summary, sub-femtosecond electron bunches with a few pC in
charge are accessible in the LWFA if electron injection is
finely controlled by 3D manipulation of the plasma bubble.
Combining a plasma density gradient with an external magnetic
field, not only modulates the bubble velocity but also the electron
longitudinal velocity. In this 3D manipulation, the increase of the bubble length along the density down-ramp increases the electron energy around the bubble rear, which results in electron injection, while an expanding hole in the bubble
rear suppresses injection. The latter is attributed to the centrifugal force, which is
proportional to the electron energy. The expanding hole will in
return reduce the electron energy around the bubble rear. As a result, prompt suppression of electron injection is achieved.
This 3D manipulation of the plasma bubble may enable realisation of
sub-femtosecond electron bunches with readily accessible
parameters both for density profiles and magnetic field strength.
Furthermore, it may be extended to generate electron bunches with
narrow energy spreads since the electrons can be properly phased
in the wake and beam loading can be compensated as long as the
electron injection is suppressed at a proper time \cite{Gonsalves2011}.

\section*{Acknowledgments}
\addcontentsline{toc}{section}{Acknowledgments}

We thank F. Y. Li for fruitful discussions. The work was supported
by the National Basic Research Program of China (Grant No.
2013CBA01504),  National Natural Science Foundation of China
(Grant Nos. 11675108, 11774227, 11721091, and 11655002), National
1000 Youth Talent Project of China, UK Engineering and Physical
Sciences Research Council (EPSRC) (Grant No. EP/N028694/1), the
European Union's Horizon 2020 research and innovation programme under grant agreements 654148 Laserlab-Europe and 653782 EuPRAXIA.


\section*{References}
\addcontentsline{toc}{section}{References}

\begin{thebibliography}{120}
\bibitem{Tajima1979} Tajima T and Dawson J M 1979 \emph{Phys. Rev. Lett.} \textbf{43} 267
\bibitem{Esarey2009} Esarey E, Schroeder C B and Leemans W P 2009 \emph{Rev. Mod. Phys.} \textbf{81} 1229
\bibitem{Pukhov2002} Pukhov A and Meyer-Ter-Vehn J 2002 \emph{Appl. Phys. B} \textbf{74} 355
\bibitem{Lu2006} Lu W, Huang C, Zhou M, Tzoufras M, Tsung F S, Mori W B and Katsouleas T 2006 \emph{Phys. Plasmas} \textbf{13} 056709
\bibitem{Yi2013} Yi S A, Khudik V, Siemon C and Shvets G 2013 \emph{Phys. Plasmas} \textbf{20} 013108
\bibitem{Jaroszynski2002} Jaroszynski D A and Vieux G 2002 \emph{Adv. Accelerator Concepts} \textbf{647} 902; Clayton C E and Muggli P Eds. 2002 \emph{in 10th Workshop on Advanced Accelerator Concepts} pp 902-913 Mandalay Beach Ca; Jaroszynski D A \emph{et al} 2006 \emph{Phil. Trans. R. Soc. A} \textbf{364} 689
\bibitem{Rousse2004} Rousse A, Phuoc K T, Shah R \emph{et al} 2004 \emph{Phys. Rev. Lett.} \textbf{93} 135005
\bibitem{Schlenvoigt2008} Schlenvoigt H P, Haupt K, Debus A \emph{et al} 2008 \emph{Nat. Phys.} \textbf{4} 130
\bibitem{Fuchs2009} Fuchs M, Weingartner R, Popp A \emph{et al} 2009 \emph{Nat. Phys.} \textbf{5} 826
\bibitem{Cipiccia2011} Cipiccia S, Islam M R, Ersfeld B \emph{et al} 2011 \emph{Nat. Phys.} \textbf{7} 867
\bibitem{Ersfeld2014} Ersfeld B, Bonifacio R, Chen S, Islam M R, Smorenburg P W and Jaroszynski D A 2014 \emph{New J. Phys.} \textbf{16} 093025
\bibitem{Chen2016} Chen M, Luo J, Li F Y, Liu F, Sheng Z M and Zhang J 2016 \emph{Light Sci. Appl.} \textbf{5} e16015
\bibitem{Faure2006} Faure J, Rechatin C, Norlin A, Lifschitz A, Glinec Y and Malka V 2006 \emph{Nature} \textbf{444} 737
\bibitem{Lundh2011} Lundh O, Lim J, Rechatin C 2011 \emph{Nat. Phys.} \textbf{7} 219
\bibitem{Kartner2016} K\"{a}rtner F X, Ahr F, Calendron A L \emph{et al} 2016 \emph{Nucl. Instrum. Methods Phys. Res. A} \textbf{829} 24
\bibitem{Dorda2016} Dorda U, Assmann R, Brinkmann R \emph{et al} 2016 \emph{Nucl. Instrum. Methods Phys. Res. A} \textbf{829} 233
\bibitem{Morimoto2018} Morimoto Y and Baum P 2018 \emph{Nat. Phys.} \textbf{14} 252
M.
\bibitem{Hassan2018} Hassan M Th 2018 \emph{J. Phys. B: At. Mol. Opt. Phys.} \textbf{51} 032005
\bibitem{Buck2011} Buck A, Nicolai M, Schmid K, Sears C M S, Savert A, Mikhailova J M, Krausz F, Kaluza M C and Veisz L 2011 \emph{Nat. Phys.} \textbf{7} 543
\bibitem{Heigoldt2015} Heigoldt M, Popp A, Khrennikov K, Wenz J, Chou S W, Karsch S, Bajlekov S I, Hooker S M and Schmidt B 2015 \emph{Phys. Rev. ST Accel. Beams} \textbf{18} 121302
\bibitem{Islam2015} Islam M R, Brunetti E, Shanks R P \emph{et al} 2015 \emph{New J. Phys.} \textbf{17} 093033
\bibitem{Li2015} Li F Y, Sheng Z M, Liu Y, Meyer-ter-Vehn J, Mori W B, Lu W and Zhang J 2013 \emph{Phys. Rev. Lett.} \textbf{110} 135002
\bibitem{Bulanov1998} Bulanov S, Naumova N, Pegoraro F and Sakai J 1998 \emph{Phys. Rev. E} \textbf{58} R5257
\bibitem{Geddes2008} Geddes C G R, Nakamura K, Plateau G R, Toth C, Cormier-Michel E, Esarey E, Schroeder C B, Cary J R and Leemans W P 2008 \emph{Phys. Rev. Lett.} \textbf{100} 215004
\bibitem{Gonsalves2011} Gonsalves A J, Nakamura K, Lin C \emph{et al} 2011 \emph{Nat. Phys.} \textbf{7} 862
\bibitem{Hansson2015} Hansson M, Aurand B, Davoine X, Ekerfelt H, Svensson K, Persson A, Wahlstr\"{o}m C G and Lundh O 2015 \emph{Phys. Rev. ST Accel. Beams} \textbf{18} 071303
\bibitem{Martinez2017} Martinez de la Ossa A, Hu Z, Streeter M J V, Mehrling T Kononenko J O, Sheeran B and Osterhoff J 2017 \emph{Phys. Rev. Accel.Beams} \textbf{20} 091301
\bibitem{Xu2017} Xu X L, Li F, An W, Dalichaouch T N, Yu P, Lu W, Joshi C and Mori W B 2017 \emph{Phys. Rev. Accel. Beams} \textbf{20} 111303
\bibitem{Tooley2017} Tooley M P, Ersfeld B, Yoffe S R, Noble A, Brunetti E, Sheng Z M, Islam M R and Jaroszynski D A 2017 \emph{Phys. Rev. Lett.} \textbf{119} 044801
\bibitem{Hosokai2006} Hosokai T, Kinoshita K, Zhidkov A Maekawa A, Yamazaki A and Uesaka M 2006 \emph{Phys. Rev. Lett.} \textbf{97} 075004; Hosokai T \emph{et al} 2010 \emph{Appl. Phys. Lett.} \textbf{96} 121501
\bibitem{Vieira2011} Vieira J, Martinss S F, Pathak V B Fonseca R A, Mori W B and Silva L O 2011 \emph{Phys. Rev. Lett.} \textbf{106} 225001; Vieira J \emph{et al} 2012 \emph{Plasma Phys. Control. Fusion} \textbf{54} 124044
\bibitem{Bulanov2013} Bulanov S V, Esirkepov T Zh, Kando M, Koga J K, Hosokai T, Zhidkov A G and Kodama R 2013 \emph{Phys. Plasmas} \textbf{20} 083113
\bibitem{Rassou2015} Rassou S, Bourdier A and Drouin M 2015 \emph{Phys. Plasmas} \textbf{22} 073104
\bibitem{Zhao2018} Zhao Q, Weng S M, Sheng Z M, Chen M, Zhang G B, Mori W B, Hidding B, Jaroszynski D A and Zhang J 2018 \emph{New J. Phys.} \textbf{20} 063031
\bibitem{Pollock2006} Pollock B B, Froula D H, Davis P F \emph{et al} 2006 \emph{Rev. Sci. Instrum.} \textbf{77} 114703
\bibitem{Fiksel2015} Fiksel G, Agliata A, Barnak D \emph{et al} 2015 \emph{Rev. Sci. Instrum.} \textbf{86} 016105
\bibitem{Shi2018} Shi Y, Vieira J, Trines R M G M, Bingham R, Shen B F and Kingham R J 2018 \emph{Phys. Rev. Lett.} \textbf{121} 145002
\bibitem{Tsakiris2000} Tsakiris G D, Gahn C and Tripathi V K 2000 Phys. Plasmas \textbf{7} 3017
\bibitem{Fonseca} Fonseca R A, Silva L O, Tsung F S \emph{et al} \emph{Lect. Not. Comput. Sci.} \textbf{2331} 342
\bibitem{Kononenko2016} Kononenko O, Lopes N, Cole J \emph{et al} 2016 \emph{Nucl. Instrum. Methods Phys. Res. Sect. A} \textbf{829} 125







\end{thebibliography}
\end{document}